\documentstyle[12pt]{article}
\renewcommand{\baselinestretch}{1.5}
\begin{document}
\begin{flushright}
MPI--PhT/95--112\\
\end{flushright}
\vspace*{5mm}
\begin{center}
{\bf Nucleon Spin and the Mixing of Axial Vector Mesons\footnote
{Supported in part by DFG--contract No.\ F412/19--3 and G.I.F. project
I--304--120--07/93}}\\
\bigskip
\end{center}
\vspace*{6mm}
\begin{center}

MICHAEL BIRKEL\\
\vspace*{3mm}
{\it Theoretical Physics, University of Oxford, \\
1 Keble Road, Oxford OX1 3NP, UK}\\
\bigskip
and\\
\bigskip
HARALD FRITZSCH\\
\vspace*{3mm}
{\it Ludwig--Maximilians--Universit\"at, Sektion Physik,\\
Theresienstrasse 37, D--80333 M\"unchen, Germany}\\
\bigskip
{\it Max--Planck--Institut f\"ur Physik, Werner--Heisenberg--Institut,\\ 
F\"ohringer Ring 6, D-80805 M\"unchen, Germany}
\end{center}
\vspace*{3mm}
\begin{center}
{\bf Abstract}\\
\end{center}
It is shown that the QCD anomaly may lead to an abnormal mixing behavior
of the axial vector mesons similar to the pseudoscalar mesons. These
mixing effects, involving a gluonic axial vector state, generate
a non-vanishing strange quark component in the nucleon. They reduce
the matrix element of the singlet axial vector in comparison to the value
obtained in a na\"\i ve quark model. The results are in agreement with the
data obtained in the polarized lepton--nucleon scattering experiments.\\
\vspace{1cm}

\noindent
November 1995
\newpage
\setcounter{page}{1}
\setcounter{equation}{0}
\renewcommand{\theequation}{1.\arabic{equation}}
\pagestyle{plain}
\section*{\large \bf 1.\ INTRODUCTION}
\bigskip
\renewcommand{\baselinestretch}{1.5}
{\small \normalsize
Deep inelastic scattering reveals that the nucleon is a complex system
consisting of an infinite number of quarks, antiquarks and gluons.
There is no doubt that the same is true for
all mesons and baryons. Nevertheless it seems that under certain circumstances
they behave as if they were composed of a single constituent quark and another
constituent antiquark or three constituent quarks. Examples are the magnetic
moments of the baryons, the spectroscopy of mesons and baryons, the 
meson--baryon couplings, the ratios of total cross sections like
$\sigma (\pi N) \, / \, \sigma (NN)$
etc.\ . Thus it seems to make sense to decompose the proton into three parts, 
into three constituent quarks called U or D. A proton would have the 
composition (UUD). The quantum numbers of the constituent quarks would provide 
the internal quantum numbers of the nucleon.\\
\\
In deep inelastic scattering one observes that a nucleon has the composition
$|u u d \bar q q...g...>$ (g: gluon, q = u,d,s); i.\ e., the quark density 
functions
(which are scale dependent) are described by a valence quark and an indefinite
number of quark--antiquark pairs. One might be tempted to identify
the valence quark, defined by the corresponding quark density function, with
a constituent quark. This identification would imply that the three--quark
picture denoted above is nothing but a very rough approximation and both
$\bar qq$--pairs and gluons need to be added to the picture. In this
case, however, one would not be able to understand why the model of a 
baryon consisting
of three constituent quarks works so well in many circumstances. It seems much
more
likely that a constituent quark is a quasiparticle 
which has a non-trivial internal structure on its own, i.\ e., consisting of a 
valence quark, of many $\bar qq$--pairs and of gluons.
Thus a constituent quark has an effective mass,
an internal size, etc. Such an interpretation of a constituent quark is not
new \cite{frigel}.
Nevertheless it is still unclear to what extent it
can be derived from the basic laws of QCD since it is deeply rooted in the
non-perturbative aspects of QCD, in particular the confinement problem.
In two dimensions the constituent quarks can be identified with specific
soliton solutions of the QCD field equations \cite{ellhan}.\\
\\
One way to gain deeper insights into the internal structure of the constituent
quarks is to consider their spin. In the
constituent quark picture it is, of course, assumed that the nucleon
spin is
provided by the combination of the spins of the three constituent quarks. If
the latter have a non-trivial internal structure, the question arises whether
also the spin structure of the constituent quarks is a complex phenomenon, as 
it seems to be the case for the nucleon, or not.
A simple model for the spin structure would be to assume that the spin of, say,
a constituent u--quark is provided by the valence u--quark inside it and the
$\bar qq$--cloud and the gluonic cloud does not contribute to the spin.
It will be one of our conclusions that this na\"\i ve picture
is not correct.\\
\\
In a na\"\i ve $SU(6)$ quark model of the baryons the spin of the proton
is composed of the spins of the three constituent quarks 
(see, e.\ g., \cite{halmar}).
The wave function in flavor and spin space is given by\\ 
\begin{equation}
| P \uparrow \rangle = \frac{1}{\sqrt{6}}
| {\rm UUD} ( 2\uparrow \uparrow \downarrow - \downarrow \uparrow \uparrow
- \uparrow \downarrow \uparrow ) \rangle .
\end{equation}
Using this wave function it is straightforward to calculate the matrix
elements of the spin operators of the various quark flavors in the proton.
If we define the quantity
\begin{equation}
\Delta Q \equiv \langle P \uparrow | \sigma^Q _z | P \uparrow \rangle ,
\end{equation}
one finds:
\begin{equation}
\Delta U = \frac{4}{3}, \quad \Delta D = - \frac{1}{3},
\quad \Delta S = 0.
\end{equation}
Since the quantity $\Delta S $ vanishes according to the wave function
given above one obtains:
\begin{equation}
\Delta U + \Delta D + \Delta S = 1 \, .
\end{equation}
As expected, the spin of the proton is carried by the spins of
the three constituent quarks.\\
\\
We should like to point out that the same calculation gives for the axial
vector coupling constant, observed in $\beta $--decay:
\begin{equation}
g_A / g_V = \Delta U - \Delta D = 5/3 \, ,
\end{equation}
while the observed value is 25\% smaller: $g_A / g_V 
\cong 1.257$ \cite{pdg}. If we interpret
this phenomenon as the result of the depolarisation of a constituent quark
by relativistic and by gluonic effects, one expects a
reduction factor of 25\% for all spin densities. Correspondingly we
would expect that the sum of the spin densities given in Eq.\ (1.4)
does not
give 1, but rather 0.75. Thus one finds that about 75\% of the
nucleon spin is carried by the spin of the constituent quarks while 25\%
are carried by orbital and gluonic effects \cite{jafman}.\\
\\
These values disagree with the measurements of the spin density functions
of the quarks carried out in the recent years \cite{exp}. In QCD
the first moment of the structure function $g_1^p$ can be expressed in terms
of the sum of the nucleon matrix elements of the axial vector currents, 
weighted by the square of the quark charges:
\begin{equation}
\int ^1_0 dxg^p_1 (x, Q^2) = \frac{1}{2} \left[
\frac{4}{9} \Delta u + \frac{1}{9} \Delta d + \frac{1}{9} \Delta s \right]
+ 0(\alpha_S / \pi) \, .
\end{equation}
\noindent
The experimental data, combined with the experimental knowledge of the
axial vector coupling constants for $\beta $--decay and hyperon decay,
give according to a recent analysis \cite{ellkar}:\\
\begin{displaymath}
\Delta u = 0.83 \pm 0.03\, , \qquad \Delta d = - 0.43 \pm 0.03\, ,
\end{displaymath}
\begin{equation}
\Delta s = - 0.10 \pm 0.03\, , \\
\end{equation}
\begin{displaymath}
\Delta \Sigma = \Delta u + \Delta d + \Delta s = 0.31 \pm 0.07 \, .
\end{displaymath}
These values are obtained at a renormalization scale $Q^2 = 10$ GeV$^2$.
They disagree with the expectations
of the na\"\i ve quark model which give in particular $\Delta S = 0$
and $\Delta \Sigma \simeq 0.75$.\\ 
\\
The sum $\Delta \Sigma$ of the three density moments 
is described by the nucleon matrix element of the
singlet axial vector current. Unlike the divergencies of the
axial vector currents of the
$SU(3)$ octet the divergence of the singlet axial vector current
does not vanish in the chiral limit, but exhibits an anomaly
because of the interaction of quarks with the gluons \cite{adler}. It has been
suggested that the anomalous behavior of the singlet axial vector current
is the reason for the observed smallness of the singlet axial vector
nucleon matrix element \cite{altros}, \cite{fri89}. Nevertheless it
remained unresolved in which way the gluon anomaly influences the spin
densities. 
On the other hand it is well known that the gluonic anomaly of the
singlet axial vector current is responsible for the anomalous mixing
behavior of the pseudoscalar mesons \cite{frimin}. It implies, for example,
that in the $SU(3)$ limit the $\eta $ meson is an $SU(3)$ octet,
while the $\eta '$ meson is an $SU(3)$ singlet. The mass difference
between these two pseudoscalar mesons is a measure for the impact
of the gluonic anomaly on the mass spectrum. In the chiral
limit $SU(3)_L \times SU(3)_R$ the eight pseudoscalar mesons act as
Nambu--Goldstone particles and are massless while the $\eta '$ meson
remains massive with a mass of order 1 GeV. The matrix elements of the
axial vector currents of the SU(3) octet exhibit a Goldstone pole
and obey a Goldberger--Treiman relation \cite{goltre}. No such relation
exists for the matrix element of the singlet axial vector current. 
This suggests that the anomalous mixing behavior of the pseudoscalar
mesons and the anomalous value for the nucleon matrix element of the 
singlet axial vector current are related \cite{fri89} and that
also the axial vector mesons might display an anomalous mixing
behavior.\\
\\
Usually it is assumed that the mixing of the axial vector mesons
is similar to the mixing of the vector mesons; i.\ e., the mass eigenstates
of the two neutral isoscalar members of the nonet segregate according to the
quark decomposition $\frac{1}{\sqrt{2}} (\bar uu + \bar dd)$ and $\bar ss$
\cite{halmar}. It is well known that this is the case for the vector mesons
being a consequence
of the Zweig rule. In the case of the axial vector mesons the
experimental situation is less clear. In this paper we should like to study
the situation of the axial vector mesons in view of the spin problem.
In particular we should like to investigate whether the anomalously
small matrix element for the singlet axial vector current could be
correlated with an anomalous mixing behavior of the axial vector mesons.\\
\\
\setcounter{section}{2}
\setcounter{equation}{0}
\renewcommand{\theequation}{2.\arabic{equation}}
\section*{\large \bf 2.\ AXIAL VECTOR MESONS AND THEIR MIXING}
\bigskip
Before turning to the axial vector mesons, we consider briefly the
pseudoscalar mesons. Here the large departure from the ideal mixing
is interpreted as the consequence of the existence of
strong transitions between the various flavor combinations \cite{frimin}, 
\cite{rujgla}.
This suggests the ansatz for the quadratic mass matrix of the pseudoscalar
mesons in the basis
$\{ | \bar uu \rangle, | \bar dd \rangle, | \bar ss \rangle\}$:
\begin{equation}
M^2_{\bar qq} =
\left( \begin{array}{ccc}
m^2_{\bar uu} + \lambda_{P} & \lambda_{P} & \lambda_{P}\\
\lambda_{P} & m^2_{\bar dd} + \lambda_{P} & \lambda_{P}\\
\lambda_{P} & \lambda_{P} & m^2_{\bar ss} + \lambda_{P}
 \end{array} \right) \, .
\end{equation}
The parameter $\lambda_{P} $ characterizes the strength of the transitions
between the various flavor eigenstates. Here $m^2_{\bar qq}$ describes
the mass of the corresponding meson in the absence of the gluonic
mixing parameter $\lambda_{P} $. In the $SU(3)$ limit the masses of
the three
flavor states are identical. They vanish in the chiral limit. It is
useful to consider the basis $\{\frac{1}{\sqrt{2}} | \bar uu - \bar dd
\rangle, \frac{1}{\sqrt{2}} | \bar uu + \bar dd \rangle, | \bar ss
\rangle\}$. In the limit of isospin symmetry the state
$\frac{1}{\sqrt{2}} | \bar uu - \bar dd \rangle$ represents the
$\pi ^0$ meson. It does not mix with the other two states because of isospin
symmetry and will be
disregarded.\\
\\
Eliminating the $\pi ^0$--state one finds in the basis
\{$\frac{1}{\sqrt{2}} | \bar uu + \bar dd \rangle,
| \bar ss \rangle$\}:
\begin{equation}
M^2 =
\left( \begin{array}{cc}
m^2_{N} + 2 \lambda_{P} & \sqrt{2}\lambda_{P}\\
\sqrt{2}\lambda_{P} & m^2_S + \lambda_{P}
 \end{array} \right) \, .
\end{equation}
Here is
\begin{equation}
m^2_N = m^2_{\bar uu} = m^2_{\bar dd} = m^2_{\pi} = (135 \mbox{ MeV})^2 \, .
\end{equation}
The parameter $m_S^2$ describes the mass splitting within the octet:
\begin{equation}
m^2_S = m^2_{\bar ss} = 2m^2_K - m^2_{\pi} = (691 \mbox{ MeV})^2 \, .
\end{equation}
The mixing angle $\Phi$ between the state 
$| S \rangle = | \bar ss \rangle$ and the isosinglet
state $| N \rangle = \frac{1}{\sqrt{2}}| \bar uu + \bar dd \rangle $ 
is given by the relation
\begin{equation}
\tan{2\Phi} = \frac{2\sqrt{2}\lambda_{P}}{\lambda_{P} + m^2_N - m^2_S}\,\, .
\end{equation}
A good description of the experimental situation is obtained for
$\lambda_{P} = 0.25 \mbox{ GeV}^2$ \cite{frimin}. Thus the system
of the pseudoscalar
mesons exhibits a strong gluonic mixing term which provides the
reason for the strong departure from the ideal mixing situation.\\
\\
In the case of the vector mesons $\omega $ and $\Phi $ one observes
a very small mixing between the strange system $| \bar ss \rangle$
and the non-strange
system $\frac{1}{\sqrt{2}}| \bar uu + \bar dd \rangle $. One finds
that the gluonic mixing parameter $\lambda_V $ for
the vector mesons is very small compared to $\lambda_{P}$:\\
$\lambda_{P} \simeq 18 \lambda _V$ \cite{sca}.\\
\\
Below we shall study  
the mixing behavior of the axial vector mesons with the quantum numbers
$J^{PC} = 1^{++}$. The spectrum of the $(1^{++})$ mesons consists of
the isovector mesons $a_1 (1260)$ and the
isoscalar mesons $f_1 (1285)$, $f_1(1420)$ and $f_1(1510)$ \cite{pdg}.
The strange mesons $K_1(1270)$ and $K_1(1400)$, which constitute an
isodoublet,
are mixtures of the corresponding $J^{PC} = (1^{++})$-- and 
$J^{PC} = (1^{+-})$--eigenstates.\\
\\
The mass of the $a_1$ is still subject to a considerable uncertainty
\cite{pdg}. To render our subsequent discussion independent of this we shall 
use a rounded mass with large error bounds thus covering the whole spectrum of
mass candidates:
\begin{equation}
m_{a_1} = (1200 \pm 110) \mbox{ MeV} \, .
\end{equation}
Similarly, for the three isoscalar $f_1 $ mesons we shall use the mass values
\begin{displaymath}
m_{f_1^{(1)}} = (1280 \pm 30) \mbox{ MeV} \, , \qquad m_{f_1^{(2)}} =
(1410 \pm 20) \mbox{ MeV} \, ,
\end{displaymath}
\begin{equation}
m_{f_1^{(3)}} = (1510 \pm 20) \mbox{ MeV} \, .
\end{equation}
Within a $SU(3)$ nonet one expects only two isoscalar mesons and we
reach the conclusion that one of the three states is not 
a $\bar qq$ meson, but rather an exotic state (gluonic meson,
multiquark state). Due to strong mixing effects one could expect that
none of the three isoscalar states
is purely of exotic nature, but all three states are mixtures
involving $\bar qq$--parts and exotic parts in their wave function.\\
\\
The mixing behavior of the strange isodoublet has been under discussion for 
a long time \cite{kintor}. In accordance with the predominating opinion, we 
shall assume a mixing angle between the $(1^{++})$-- and 
$(1^{+-})$--eigenstates of 
45$^{\circ }$. This mixing pattern implies that the mass eigenvalue of the
strange isodoublet in the $(1^{++})$--octet is determined by
the average value of the $K_1 (1270)$-- and $K_1 (1400)$--states
(in [mass]$^2$):
\begin{equation}
m_{K_1} = (1340 \pm 30) \mbox{ MeV} \, .
\end{equation}
In analogy to the pseudoscalar mesons we shall investigate the mixing
pattern of the axial vector mesons. In order to accommodate an exotic
state, we shall extend the mass matrix given in Eq.\ (2.1).
To be more specific, we shall assume that the exotic
configuration in the axial vector channel is of gluonic nature
denoted by $| G \rangle$.
The mass matrix of the axial vector mesons has then the form: \\
\begin{equation}
M^2_0 = 
\left( \begin{array}{cccc}
m^2_N + \lambda & \lambda & \lambda & \kappa\\
\lambda & m^2_N + \lambda & \lambda & \kappa\\
\lambda & \lambda & m^2_S + \lambda & \kappa\\
\kappa & \kappa & \kappa & m^2_G
 \end{array} \right)
\end{equation}
\bigskip
\begin{displaymath}
m^2_N = m^2_{a_1} \, , \quad \, m^2_S = 2m^2_{K_1} - m^2_{a_1} \, \, . 
\end{displaymath}
Again, the parameter $\lambda $ describes the strength of the mixing
between the various $\bar qq$--configurations while the parameter
$\kappa $ describes the transition between a $\bar qq$--configuration and 
the gluonic configuration. Of course, in the special case
$\kappa = 0$ the mass $m_G$ corresponds to the mass of the physical
gluonic state.
In Eq.\ (2.9) we did not denote a contribution of $\kappa $
in the 44--matrix element since it is included in ${m_G}^2$.\\
\\
We shall denote the coefficients of the various $f_1$ mesons
in the basis\\ 
\{$| N \rangle = \frac{1}{\sqrt{2}} | \bar uu + \bar dd \rangle \, ,
| S \rangle = | \bar ss \rangle , | G \rangle$\}
by $x_i$, $y_i$, $z_i$:\\
\begin{equation}
| f^{(i)}_1 \rangle
= x_i | N \rangle + y_i | S \rangle + z_i | G \rangle.
\end{equation}
In order to calculate these coefficients, we shall proceed
in an analogous way as in the case of the pseudoscalar mesons.
After excluding the isotriplet configuration, which must remain unmixed
because of isospin symmetry, we obtain from Eq.\ (2.9) \\
\begin{equation}
M^2 = 
\left( \begin{array}{ccc}
m^2_N + 2\lambda & \sqrt{2}\lambda & \sqrt{2}\kappa\\
\sqrt{2}\lambda & m^2_S + \lambda & \kappa\\
\sqrt{2}\kappa & \kappa & m^2_G
\end{array} \right)\, .
\end{equation}
The mass matrix of the mass eigenvalues is denoted by \\
\begin{equation}
M^2_m = 
\left( \begin{array}{ccc}
m^2_{f^{(1)}_1} & 0 & 0\\
0 & m^2_{f^{(2)}_1} & 0\\
0 & 0 & m^2_{f^{(3)}_1}
\end{array} \right) \, .
\end{equation}
Let us define the quantities $l_i$:\\
\begin{equation}
 \begin{array}{lll}
l_1 & \equiv & \mbox{tr } M^2 = \mbox{tr } M^2_m,\\
l_2 & \equiv & \frac{1}{2} \left[ (\mbox{tr } M^2)^2 - \mbox{tr}
(M^2)^2 \right]\\
& = & m^2_{f^{(1)}_1} m^2_{f^{(2)}_1} + m^2_{f^{(1)}_1} m^2_{f^{(3)}_1}
+ m^2_{f^{(2)}_1} m^2_{f^{(3)}_1},\\
l_3 & \equiv & \mbox{det } M^2 = \mbox{det } M^2_m.
\end{array}
\end{equation}
For the parameters $\lambda$, $\kappa$ and $m_G$ we then obtain:\\
\begin{equation}
\lambda = a/b, \quad \kappa
= \sqrt{(c \lambda^2 + d\lambda + e)/(m^2_N + 2m^2_S)}\, , \quad 
m_G = \sqrt{l_1 - m^2_N - m^2_S - 3 \lambda}\, ,
\end{equation}
\newpage
\noindent
where \\
\begin{equation}
\begin{array}{lll}
a & \equiv & l_1(m^2_N + m^2_S) - m^4_N - m^4_S -l_2 -m^2_Nm^2_S +\\
& & + 3(m^4_Sm^2_N + m^4_Nm^2_S + l_3 - m^2_Nm^2_Sl_1)/(m^2_N + 2m^2_S),\\
b & \equiv & 2(m^2_N - m^2_S)^2 / (m^2_N + 2m^2_S),\\
c & \equiv & -3(m^2_N + 2m^2_S),\\
d & \equiv & m^2_N(l_1 - m^2_N) + m^2_S (2l_1 - 6m^2_N -2m^2_S),\\
e & \equiv & m^2_N(l_1m^2_S - m^2_N m^2_S - m^4_S) - l_3.
\end{array}
\end{equation}
The unitary matrix
\begin{equation}
U = 
\left( \begin{array}{ccc}
x_1 & y_1 & z_1\\
x_2 & y_2 & z_2\\
x_3 & y_3 & z_3
\end{array} \right)
\end{equation}
describes the transformation between the basis
\{$| N\rangle, | S \rangle, | G \rangle $\}
and the basis\\
\{$| f^{(1)}_1 \rangle, | f^{(2)}_1\rangle, | f^{(3)}_1\rangle$\}.
The equation $U M^2 U^{-1} = M^2_m$ then leads to the mixing parameters \\
\begin{equation}
\begin{array}{lll}
x_i & = & \sqrt{2}\kappa\left( m^2_S - m^2_{f^{(i)}_1}\right) c_i,\\
y_i & = & \kappa\left( m^2_N - m^2_{f^{(i)}_1}\right) c_i,\\
z_i & = & \left[ 2\lambda ^2 - \left( m^2_N
+ 2\lambda - m^2_{f^{(i)}_1} \right)
\left( m^2_S + \lambda - m^2_{f^{(i)}_1}\right) \right] c_i \, .
\end{array}
\end{equation}
The normalization constants $c_i$ are given by \\
\begin{eqnarray}
c_i & = & \left\{ 2 \kappa ^2 \left( m^2_S - m^2_{f^{(i)}_1} \right)^2 + 
\kappa ^2 \left( m^2_N - m^2_{f^{(i)}_1} \right)^2 + \right. \nonumber \\
& + & \left. \left[ 2 \lambda ^2
-\left( m^2_N + 2\lambda - m^2_{f^{(i)}_1} \right) \left( m^2_S
+ \lambda - m^2_{f^{(i)}_1} \right) \right] ^2 \right\} ^{-\frac{1}{2}}.
\end{eqnarray}
\\
Using the meson masses mentioned previously we calculate the numerical 
results for the coefficients
$x_{i}$ etc.\ as shown in Table 1.
The uncertainties in the coefficients are relatively large because of the fact
that the meson masses are not precisely known.\\
\newpage
\noindent
\begin{table}[htb]
\caption[koeerg]{\small Coefficients of the 
$f_{1}$ mesons.}\label{koeerg}
   \begin{center}
      \begin{tabular}{|c|c|c|c|}
      \hline
      & $f_{1}^{(1)}$ & $f_{1}^{(2)}$ & $f_{1}^{(3)}$ \\
      \hline
      $x_{i}$ & $0.93\pm 0.05$ & $0.09\pm 0.17$ & $\!\!\!-0.30\pm 0.11$ \\
      \hline
      $y_{i}$ & $\!\!\!-0.20\pm 0.14$ & $\!\!\!-0.55\pm 0.23$ &
$\!\!\!-0.74\pm 0.19$ \\
      \hline
      $z_{i}$ & $\!\!\!-0.25\pm 0.13$ & $0.76\pm 0.19$ & $\!\!\!-0.51\pm
0.23$ \\     
      \hline
      $\lambda\, [${\small GeV}$^{2}]$ & \multicolumn{3}{c|}{$0.10\pm 0.04$}
\\
      \hline
      $\kappa\, [${\small GeV}$^{2}]$ & \multicolumn{3}{c|}{$0.10\pm 0.03$}
\\
      \hline
      $m_{G}\, [${\small MeV}$]$ & \multicolumn{3}{c|}{$1432\pm 38$} \\
      \hline
      \end{tabular}
   \end{center}
\end{table}
\\
\\
We find a mixing behavior of the $f_{1}$ mesons 
which is quite different from the mixing behavior of the vector mesons.
Such a conclusion has also been reached after an analysis of the
radiative decays $J/\psi \rightarrow \gamma f_1(1285)$, observed by the
Mark III
collaboration \cite{bol}. Relatively large mixing exists between the 
$|N\rangle$--, $|S\rangle$-- and $|G\rangle$--states. This corresponds
to a relatively
large violation of the OZI--rule. Let us, as an illustrative example, consider
the coefficients for the masses 
$m_{a_1} = 1215 \mbox{ MeV}$, $m_{K_1} = 1320 \mbox{ MeV}$, 
$m_{f_1^{(1)}} = 1275 \mbox{ MeV}$, $m_{f_1^{(2)}} = 1390 \mbox{ MeV}$, 
$m_{f_1^{(3)}} = 1540 \mbox{ MeV}$ as given in Eq.\ (2.19). The 
coefficients are in agreement with the observations of the various decays
of the $f_1$ mesons \cite{pdg}. For instance, the $f_1^{(2)}$ meson decays 
dominantly into $K \bar K \pi$ and the meson $f_1^{(3)}$ into 
$K \bar K^*(892) + c.c.\, $. This is expected since according
to Eq.\ (2.19) both states have a relatively large 
$\bar s  s$--component.\\
\begin{eqnarray}
   |f^{(1)}_{1}\rangle & \simeq & 0.89\, |\frac{1}{\sqrt{2}}(\bar u u + \bar
d d)\rangle - 0.25\, |\bar s s\rangle - 0.38\, |G\rangle\, , \nonumber\\
   |f^{(2)}_{1}\rangle & \simeq & 0.15\, |\frac{1}{\sqrt{2}}(\bar u u + \bar
d d)\rangle - 0.63\, |\bar s s\rangle + 0.76\, |G\rangle\, , \\
   |f^{(3)}_{1}\rangle & \simeq & \!\!\!-0.42\, |\frac{1}{\sqrt{2}}(\bar u u
+ \bar d d)\rangle - 0.74\, |\bar s s\rangle - 0.52\, |G\rangle\, .\nonumber
\end{eqnarray}
\\
According to Eq.\ (2.19) the meson $f_1^{(2)}$ has a probability
of 58\% to be a gluonic meson while this probability is reduced
for the $f_1^{(1)}$ meson to 14 \% and for the $f_1^{(3)}$ meson to 27 \%.\\
\\
The strong mixing in the axial vector meson channel which we obtain
reminds us of the corresponding situation for the pseudoscalar mesons.
Since in the latter case the $U(1)$ anomaly is responsible for the
large mixing behavior we conclude that the gluonic anomaly
also influences the mixing pattern of the axial vector mesons.
\\
Using $SU(3)$ symmetry, it is also useful to describe the meson mixing
in terms of the basis \{$\frac{1}{\sqrt{6}} | \bar uu + \bar dd - 2\bar 
ss \rangle, \frac{1}{\sqrt{3}} | \bar uu + \bar dd + \bar ss \rangle ,
| G \rangle$\}. The coefficients of the 
$f_1$ mesons in this basis are \\
\begin{eqnarray}\label{xtilde}
     \tilde x_{i} & = & (x_{i} - \sqrt{2}y_{i})/\sqrt{3} \, , \nonumber\\
     \tilde y_{i} & = & (\sqrt{2}x_{i} + y_{i})/\sqrt{3} \, , \\
     \tilde z_{i} & = & z_{i} \, . \nonumber
\end{eqnarray}
\bigskip
\setcounter{section}{3}
\setcounter{equation}{0}
\renewcommand{\theequation}{3.\arabic{equation}}
\section*{\large \bf 3.\ \parbox[t]{14cm}
{MATRIX ELEMENTS OF THE AXIAL VECTOR CURRENTS}}
\bigskip
In this section we shall calculate the proton matrix elements
of the various axial vector
currents using the idea of axial vector dominance. In analogy to the case of
vector meson dominance we shall assume that the matrix elements of the
axial vector currents are dominated by the contribution of the lowest lying
axial vector mesons. Thus we obtain the relation
\begin{equation}
\langle p | \bar q \gamma_{\mu } \gamma_5 q | p \rangle =
\left. \sum _A \frac{\langle 0 | \bar q \gamma _{\mu } \gamma _5 q |
A \rangle \langle Ap | p \rangle}{m^2_A - k^2} \right| _{k=0} \, .
\end{equation}
Here the matrix element
$\langle 0 | \bar q \gamma_{\mu } \gamma_5 q |
A \rangle$
denotes the transition element of the axial vector current between the
vacuum
state and the corresponding axial vector meson while the second factor
$\langle Ap | p \rangle$ describes the coupling of the
axial vector meson to the proton. The four--momentum transfer is denoted by
$k$. The summation in Eq.\ (3.1) is carried out over all axial vector mesons
with the quantum numbers
$J^{PC} = 1^{++}$ which can couple to the proton, i.\ e.,
the mesons $a_1, f_1^{(1)}, f_1^{(2)}$ and
$f_1 ^{(3)}$. Once these matrix elements are known we can
calculate $\Delta q$. \\
\\
First we consider the matrix element of the third component of the
isovector
\begin{equation}
\sqrt{2}\langle 0 | A_{\mu }^3 | a_1 \rangle = \langle 0 | 
\frac{1}{\sqrt{2}} (\bar u \gamma _{\mu } \gamma _5 u - \bar d
\gamma_{\mu } \gamma _5 d) | \frac{1}{\sqrt{2}}
(\bar uu - \bar dd) \rangle \, .
\end{equation}
The decay constant for the $a_1$ is defined by
\begin{equation}
i\varepsilon^r_{\mu} f_{a_1} \equiv \sqrt{2} \langle 0 | A_\mu ^3 |
a_1 \rangle
\end{equation}
($\varepsilon_{\mu }^r$: polarisation vector for the $a_1$ meson).
This equation can be rewritten as follows:
\begin{equation}
i \varepsilon^r_{\mu } \cdot 2f_{a_1} = \langle 0 | \bar u \gamma _{\mu }
\gamma _5 u | (\bar uu)_{a_1} \rangle + \langle 0 | \bar d
\gamma _{\mu } \gamma _5 d | (\bar dd)_{a_1} \rangle \, .
\end{equation}
Here we denote by $(\bar uu)_{a_1}$ the $\bar uu$--part of the $a_1$ meson
etc. Using $SU(3)$ symmetry we can define the decay constants of the
axial vector meson of quark composition ($\bar qq$) by
\begin{equation}
i \varepsilon ^r_{\mu } f_A = \langle 0 | \bar q \gamma _{\mu } 
\gamma _5 q | (\bar q q)_A \rangle \, .
\end{equation}
The matrix element for the coupling of the meson with the nucleon
$\langle a_1 \, p | p \rangle $ can be written in terms of the
Dirac wave functions:
\begin{equation}
\langle a_1 \, p | p \rangle
= ig_{a_1pp} \bar u (p) \gamma ^{\nu } \gamma _5 u (p)
\varepsilon ^r _{\nu } \, ,
\end{equation}
where $g_{a_1 pp}$ denotes the coupling constant of the meson to the
nucleon. \\
\\
If we consider one of the $f_1$ mesons, we must take into account the
mixing of these neutral mesons among each other. In what follows we shall
denote the octet state by \\
$| f_8 \rangle  = \frac{1}{\sqrt{6}} | \bar uu + \bar dd -
2 \bar ss \rangle $ and the singlet state by 
$| f_0 \rangle = \frac{1}{\sqrt{3}} | \bar uu + \bar dd +
\bar ss \rangle $. For the corresponding decay constants we shall assume
$f_{f_0}=f_{f_8}=f_{a_1}$ and for the coupling constants 
$g_{f_0 pp}=\sqrt{2}g_{f_8 pp}$.  \\
\\
Taking into account our description for the mixing of the mesons, we obtain
\begin{equation}
\langle 0 | \bar q \gamma_{\mu } \gamma _5 q | f_1^{(i)} \rangle
= \tilde x_i \langle 0 | \bar q \gamma _{\mu } \gamma _5 q
| f_8 \rangle + \tilde y_i \langle 0 | \bar q \gamma _{\mu } \gamma _5 q
| f_0 \rangle + \tilde z _i \langle 0 | \bar q \gamma _{\mu }
\gamma _5 q | G \rangle \, .
\end{equation}
The last term in this equation refers to the transition of the axial
vector current to a gluonic state. We shall assume in accordance
with the Zweig rule that the transition of the axial vector current, which
is a quark bilinear, to the gluonic state vanishes. \\
\\
Subsequently we discuss the numerical results.
We determine the decay constant $f_{a_1}$ in considering the decay
$\tau ^- \rightarrow a_1^- + \nu_{\tau }$ \cite{tsai}.
We find
\begin{equation}
f_{a_1} = (0.19 \pm 0.03) \mbox{ GeV}^2 \, .
\end{equation}

The coupling constant $g_{a_1pp}$ is
difficult to determine directly.
However, we can relate it to the observed axial vector coupling constant
using axial vector dominance:
\begin{equation}
\frac{1}{6} \frac{g_A}{g_V} = \frac{1}{6} (\Delta u - \Delta d)
= \frac{f_{a_1} g_{a_1pp}}{3 \sqrt{2}m^2_{a_1}} \, .
\end{equation}
Using this constraint, we can determine the $a_1$--nucleon coupling constant:
\begin{equation}
g_{a_1pp} = 6.7 \pm 1.0 \, .
\end{equation}
The coupling constants of the various axial vector mesons to the
baryons can be described in terms of reduced elements $F$ and $D$,
using $SU(3)$ symmetry (see, e.\ g., \cite{bern}). One has
\begin{equation}
\begin{array}{l@{\:=\:}l@{\:=\:}l}
g_{a_1^- \bar n p} & g_{a_1^+ \bar p n} & \sqrt{2}(F + D) \, , \\
g_{a_1pp} & - g_{a_1nn} & (F + D) \, , \\
g_{f_8pp} & g_{f_8nn} & \sqrt{3}F - D / \sqrt{3} \nonumber \, .
\end{array}
\end{equation}
Unlike the corresponding $SU(3)$ matrix elements for the axial vector
currents, the reduced matrix elements for the coupling of the axial vector
mesons to the baryons are not known. We can estimate them by using the same
relation between $F$ and $D$ as inferred from the measurements for the 
corresponding reduced matrix elements for the axial vector currents \cite{pdg}:
\begin{equation}
\frac{F}{D} = 0.575 \pm 0.016 \, .
\end{equation}
In this case we get
\begin{equation}
g_{f_8pp} = 1.8 \pm 0.4 \, , \hspace{0.5cm} g_{f_0pp} = 2.5 \pm 0.5 \, .
\end{equation}
\\
Subsequently we shall calculate $\Delta q$, the singlet sum $\Delta \Sigma$
and the first moment of the structure function $g_1^p$ taking into account
the mixing behavior of the axial vector mesons\footnote[1]{Due to the anomalous
dimension for the singlet axial vector current, the density moments are scale
dependent. This scale dependence, however, is relatively weak and therefore
does not affect our results within the error bounds 
(see, e.\ g., \cite{jaff}).}. As an illustration
we shall first consider a hypothetical situation in which $SU(3)$
is considered to be exact and the three neutral axial vector mesons
are denoted by $| a_1 \rangle, | f_8 \rangle $
and $| f_0 \rangle$ ($a_1$ and $f_8$ are degenerate in the symmetry
limit). One finds for the density moments:
\begin{eqnarray}
\Delta u^{(1)} & = & \frac{1}{\sqrt{6}}\frac{f_{f_8}g_{f_8pp}}{m^2_{f_8}} + 
\frac{1}{\sqrt{3}}\frac{f_{f_0}g_{f_0pp}}{m^2_{f_0}}
+ \frac{1}{\sqrt{2}}\frac{f_{a_1}g_{a_1pp}}{m^2_{a_1}} \nonumber \, , \\
\Delta d^{(1)} & = & \frac{1}{\sqrt{6}}\frac{f_{f_8}g_{f_8pp}}{m^2_{f_8}} + 
\frac{1}{\sqrt{3}}\frac{f_{f_0}g_{f_0pp}}{m^2_{f_0}}
- \frac{1}{\sqrt{2}}\frac{f_{a_1}g_{a_1pp}}{m^2_{a_1}} \, , \\
\Delta s^{(1)} & = & -\frac{2}{\sqrt{6}}\frac{f_{f_8}g_{f_8pp}}{m^2_{f_8}} + 
\frac{1}{\sqrt{3}}\frac{f_{f_0}g_{f_0pp}}{m^2_{f_0}}  \nonumber \, .
\end{eqnarray}
It is easy to see that the 
strange density moment $\Delta s$ vanishes
in the symmetry limit if
$f_0$ and $f_8$ are degenerate. If the singlet state is heavier than
the octet state, a non-zero contribution for $\Delta s$ is generated
which is expected to be negative.
In the formal limit $m_{f_0} \rightarrow \infty$ we obtain
\begin{displaymath}
\Delta u^{(1)} = 0.77 \pm 0.18, \qquad
\Delta d^{(1)} = -0.56 \pm 0.13, \qquad \Delta s^{(1)} = -0.20 \pm 0.05,
\end{displaymath}
\begin{equation}
\Delta \Sigma ^{(1)} = 0, \qquad
\int ^1_0 dxg_1 ^{p,(1)}(x) = 0.13 \pm 0.03 \, , 
\end{equation}
results which exhibit the correct qualitative features of the 
experimental data discussed previously.
The singlet sum $\Delta \Sigma$ vanishes because of the vanishing of the
contribution of the singlet axial vector meson $f_0$ in the limit
$m_{f_0} \rightarrow \infty $.\\
\\
It is also useful to consider the following case with the finite mass
$m_{f_0}$. If we take as an illustrative example $m_{f_0} = m_{f_1^{(3)}}$,
we get in the case of $SU(3)$ symmetry
\begin{displaymath}
\Delta u^{(1)} = 0.89 \pm 0.22, \quad
\Delta d^{(1)} = -0.43 \pm 0.10, \quad
\Delta s^{(1)} = -0.07 \pm 0.03,
\end{displaymath}
\begin{equation}
\Delta \Sigma^{(1)} = 0.39 \pm 0.13, \quad
\int ^1_0 dxg_1 ^{p,(1)}(x) = 0.17 \pm 0.04 \, .
\end{equation}
As compared to the previous case, the $|\Delta s|$ is reduced
by more than a factor of 2 while $\Delta \Sigma $ increases considerably.
Of course, comparing these values with the experimental
data is not useful at this stage since $SU(3)$ breaking has not yet been taken
into account. We proceed to do so by replacing the states
$| f_0 \rangle $ and $| f_8 \rangle$ by the states
$| f_1^{(i)} \rangle$. In this case the moments
of the density functions become
\begin{eqnarray}
\Delta u^{(2)} & = & \frac{f_{f_8}g_{f_8pp}}{\sqrt{6}} \sum_{i=1}^3
\frac{1}{m^2_{f_1^{(i)}}} (\tilde x_i^2 + 2 \tilde y_i^2 + 2 \sqrt{2}
\tilde x_i \tilde y_i) + \frac{f_{a_1}g_{a_1pp}}{\sqrt{2}m^2_{a_1}}
\nonumber \, , \\
\Delta d^{(2)} & = & \frac{f_{f_8}g_{f_8pp}}{\sqrt{6}} \sum_{i=1}^3
\frac{1}{m^2_{f_1^{(i)}}} (\tilde x_i^2 + 2 \tilde y_i^2 + 2 \sqrt{2}
\tilde x_i \tilde y_i) - \frac{f_{a_1}g_{a_1pp}}{\sqrt{2}m^2_{a_1}} \, , \\
\Delta s^{(2)} & = & -\frac{2}{\sqrt{6}}f_{f_8}g_{f_8pp} \sum_{i=1}^3
\frac{1}{m^2_{f_1^{(i)}}} (\tilde x_i^2 - \tilde y_i^2 + 
\tilde x_i \tilde y_i / \sqrt{2}) \nonumber \, .
\end{eqnarray}
As expected we obtain the same contributions as obtained previously
in the $a_1$--channel while the contributions of the $f_1$ mesons
are modified by the mixing terms. In the special case
$\tilde x_i = \delta_{1i},\, \, \tilde y_i = \delta_{2i}$ we can reconstruct
the case discussed above.
Using the numerical results from section 2 we find:
\begin{displaymath}
\Delta u^{(2)} = 0.92 \pm 0.21, \quad
\Delta d^{(2)} = -0.38 \pm 0.09, \quad
\Delta s^{(2)} = -0.02 \pm 0.01,
\end{displaymath}
\begin{equation}
\Delta \Sigma ^{(2)} = 0.52 \pm 0.13, \quad
\int ^1_0 dxg_1 ^{p,(2)}(x) = 0.18 \pm 0.04 \, .
\end{equation}
It is typical for this case that one obtains a rather small contribution
$|\Delta s|$ and a relatively large value of $\Delta \Sigma $.\\
\\
Thus far we have not taken into account the direct
coupling of the gluonic state to the nucleon in assuming
$\langle Gp | p \rangle = 0$. In view of the fact that gluons contribute
a large part of the momentum of a fast moving nucleon such a constraint
is highly unrealistic. As soon as a direct coupling of the nucleon to the
gluonic state is introduced by setting $\langle Gp | p \rangle \neq 0$,
one finds that the density moments have the form
\begin{eqnarray}
\Delta u^{(3)} & = & \Delta u^{(2)} + \frac{f_{f_8}g_{Gpp}}{\sqrt{6}}
\sum_{i=1}^3 \frac{1}{m^2_{f_1^{(i)}}} \tilde z_i(\tilde x_i + \sqrt{2}
\tilde y_i)\, , \nonumber \\
\Delta d^{(3)} & = & \Delta d^{(2)} + \frac{f_{f_8}g_{Gpp}}{\sqrt{6}}
\sum_{i=1}^3
\frac{1}{m^2_{f_1^{(i)}}} \tilde z_i (\tilde x_i + \sqrt{2} \tilde y_i)\, ,\\
\Delta s^{(3)} & = & \Delta s^{(2)} + \frac{f_{f_8}g_{Gpp}}{\sqrt{6}}
\sum_{i=1}^3
\frac{1}{m^2_{f_1^{(i)}}} \tilde z_i(-2 \tilde x_i + \sqrt{2} \tilde y_i)
\nonumber \, . 
\end{eqnarray}
Of course, the coupling constant $g_{Gpp}$ is not known. We shall treat it
as a free parameter. As an example we use $g_{Gpp} = 19:$
\begin{displaymath}
\Delta u^{(3)} = 0.83 \pm 0.20, \quad
\Delta d^{(3)} = -0.48 \pm 0.11, \quad
\Delta s^{(3)} = -0.10 \pm 0.03,
\end{displaymath}
\begin{equation}
\Delta \Sigma^{(3)} = 0.25 \pm 0.15, \quad
\int ^1 _0 dxg_1^{p, (3)}(x) = 0.15 \pm 0.04 \, .
\end{equation}
As one can see, we find a relatively good agreement between observation
and the results of axial vector meson dominance, provided the mixing and
a relatively large non-vanishing 
coupling of the gluonic state of the nucleon is taken into account.
In particular
we find a negative contribution to $\Delta s$. The sign of the strange
density moment is determined by the same mechanism as in the hypothetical
case discussed previously where it arises because of the non-degeneracy of
the singlet and the octet states. Since the octet is lower in mass than
the singlet,
the negative sign of the $\bar ss$--component in the octet state leads to
the negative sign of the strange density moment. Of course, the opposite
effect would be expected for the unrealistic case where the singlet
state has a smaller mass than the octet
state. A good description of the experimental situation is obtained, if the
mixing among the neutral axial vector mesons is described according to
the mixing scheme discussed above and if the nucleon has a fairly strong
coupling to the gluonic state.\\
\\
It is also instructive to observe that in the hypothetical limit in which
no mixing between the various $\bar qq$--axial vector mesons takes place
($\lambda = \kappa = 0$) one has
\begin{equation}
\tilde x_1 = \tilde y_2 = \frac{1}{\sqrt{3}}, \quad \tilde y_1
= - \tilde x_2
= \sqrt{\frac{2}{3}}\, ,
\end{equation}
\begin{displaymath}
\tilde x_3 = \tilde y_3 = \tilde z_1 = \tilde z_2 = 0, \quad \tilde z_3 =
1 \, .
\end{displaymath}
The masses of the $f_1$ mesons are given by  
$m^2_{f_1^{(1)}} = m^2_{a_1}$, $m^2_{f_1^{(2)}} = m^2_S$ and
$m^2_{f_1^{(3)}} = m^2_G$. This is, of course, the limiting case
in which the Ellis--Jaffe sum rule \cite{elljaf} is valid and now we have
$\Delta s = 0$.
In particular we can see that there is a direct
link between the vanishing of the mixing parameters $\lambda $ and
$\kappa $ and the vanishing of the strange density moment in the
nucleon. At the same time, we find that the singlet sum $\Delta \Sigma 
\rightarrow 1$ as $\lambda, \kappa \rightarrow 0$. 
In the case of the pseudoscalar mesons the vanishing of the
mixing parameter $\lambda $ implies that the gluonic anomaly is not
present, and there is a degeneracy between the singlet and the octet
pseudoscalar mesons. As we have suggested, a similar phenomenon is
supposed to occur in the axial vector meson channel. Thus, we can make the
gluonic anomaly responsible for a non-vanishing strange quark moment of the
nucleon.
At present it is
not clear whether this phenomenon, which implies a large violation
of the Zweig rule in the $(1^{++})$--channel, is directly related to
the $U(1)$ problem. 
At this point it is interesting to consider recent analyses of the 
singlet
axial channel based on the sum rule technique. The authors
of reference \cite{iofkho} aim to evaluate $\Delta \Sigma$ in a way
similar to the calculation of the octet axial constant but by taking 
into account the
presence of the anomaly in the singlet axial channel. This
analysis leads to the same conclusion as the one above, namely
that the singlet axial channel 
is
qualitatively similar to the pseudoscalar channel and
differs much
from the corresponding situation in the vector channel.
In a more recent study \cite{nashve}, the matrix elements arising from
the operator product expansion for the deep 
inelastic scattering amplitude are factorised into composite operator
propagators and proper vertex functions. 
Whereas the vertex is evaluated 
according to the OZI--rule, the propagator, being RG non--invariant,
is computed using QCD spectral sum rules. 
As a result, this paper also obtains $\Delta \Sigma$ 
in agreement with the experimental data and finds its suppression
to be a consequence of the anomaly.
In a different approach \cite{belter}, $\Delta \Sigma$ is 
calculated in the framework of QCD sum rules using
an interpolating nucleon current which explicitely contains the
gluonic degrees of freedom.
Again, $\Delta \Sigma$ is obtained in agreement with 
observation and the conclusions are the same as in reference
\cite{nashve}.\\
\\
Finally we should like to emphasize that the successful description
of the axial vector meson situation by our mixing scheme implies that the
three observed neutral isosinglet axial vector mesons $f_1$ are indeed
superpositions of $\bar uu /\bar dd$, $\bar ss$ and gluonic states. As
expected, the mixing among the three states is large, i.\ e., none of
the states can be considered to be a pure $\bar qq$ or pure gluonic
state. While the results for the mixing parameters estimated by us
are subject to a large uncertainty, it is 
important to note that within our approach the gluonic anomaly
manifests itself also in the $(1^{++})$--channel. The existence of
gluonic states in the $(1^{++})$--channel, their mixing with the
$\bar qq$--states and the problem of the nucleon spin are intimately
related. This strengthens the idea that the problem of the nucleon
spin is intrinsically related to non-perturbative aspects of chiral
QCD dynamics.\\
\newpage
\setcounter{section}{4}
\section*{\large \bf 4.\ CONCLUSIONS}
\bigskip
In this paper we have described a mixing pattern within the axial vector
meson channel by similar methods as used to describe the mixing in the
pseudoscalar meson channel where the gluonic anomaly is operating.
We suggest that in the case of the axial vector currents the behavior
of the longitudinal part and of the transverse part of the current
matrix elements is qualitatively similar, implying a large violation
of the Zweig rule also in the axial vector channel. We have shown that
a consistent picture emerges provided there are three neutral isosinglet
axial vector states, in accordance with the experimental observation.
These three states are superpositions of $\bar uu / \bar dd, \bar ss$
and gluonic states. The mixing among these states is large. It leads to
a non-vanishing contribution of the strange density function in the
nucleon. The sign of the strange density function is negative because of the
fact that the isosinglet state with the smallest mass contributing to the
corresponding matrix element is close to an $SU(3)$ octet, i.\ e., to
the state $\frac{1}{\sqrt{6}} (\bar uu + \bar dd - 2 \bar ss)$. The
mixing which is essential in order to arrive at a correct description
of the observed spectrum leads at the same time to a 
non-vanishing polarized strange quark density and to a reduction of
the singlet sum $\Delta \Sigma $.
Since the mixing reflects
directly a non-perturbative feature of the gluonic anomaly, we find 
a direct link between the gluonic anomaly as a non-perturbative
feature of the dynamics of the nucleon and the nucleon spin. It remains
to be seen whether the large mixing in the axial vector meson channel
suggested here can indeed be obtained in taking into account the
non-perturbative features of QCD, \mbox{e.\ g.\ }in the lattice approach
or by considering QCD sum rules.
For example, we would expect that the two--point function
$\langle 0 | \bar u(x) \gamma _{\mu } \gamma _5 u(x) \bar d(y)
\gamma _{\nu} \gamma _5 d(y) | 0 \rangle $,
which vanishes in the absence of a gluonic interaction, receives strong
contributions not only in the longitudinal part (because of the pseudoscalar
gluonic anomaly), but also in the transverse part. These effects should
be investigated in more detail using perturbative techniques \cite{doming}.\\
\vspace{0.3cm}
\newpage
\noindent
{\large \bf ACKNOWLEDGMENTS}\\
We would like to thank Profs.\ C.\ Dominguez, G.\ Karl, M.\ Karliner,
A.\ Khodjamirian and R.\ R\"uckl as well as M.\ Hutter for helpful discussions.


\newpage
\begin{thebibliography}{99}
\bibitem{frigel}
   H.\ Fritzsch and M.\ Gell-Mann, Proc. of XVI. International Conference on\\
   High Energy Physics, Batavia, edited by J.\ D.\ Jackson, A.\ Roberts, \\
   and R.\ Donaldson (National Accelerator Laboratory, Chicago 1972), Vol.\ 2,
   p.\ 135.

\bibitem{ellhan}
   J.\ Ellis, Y.\ Frishman, A.\ Hanany and M.\ Karliner,
   Nucl.\ Phys.\ {\bf B382} (1992) 189.

\bibitem{halmar} 
   J.\ J.\ Kokkedee, {\it The Quark Model} (W.\ A.\ Benjamin,
   New York, 1969);\\
   F.\ E.\ Close, {\it An Introduction to Quarks and Partons}
   (Acad.\ Press, London, 1979);\\
   F.\ Halzen and A.\ D.\ Martin, {\it Quarks and Leptons}
   (John Wiley \& Sons, 1984).

\bibitem{pdg}
   Review of Particle Properties, Phys.\ Rev.\ {\bf D50} (1994) 1173.

\bibitem{jafman}
   H. Fritzsch, Mod.\ Phys.\ Lett.\ {\bf A5} (1990) 625; \\
   R.\ L.\ Jaffe and A.\ Manohar, Nucl.\ Phys.\ {\bf B337} (1990) 509.

\bibitem{exp}
   J.\ Ashman {\it et al.} (EMC), Phys.\ Lett.\ {\bf B206} (1988) 364; 
   Nucl.\ Phys.\ {\bf B328} (1989) 1; \\
   P.\ L.\ Anthony {\it et al.} (E142), Phys.\ Rev.\ Lett.\ {\bf 71} (1993)
   959;\\
   B.\ Adeva {\it et al.} (SMC), Phys.\ Lett.\ {\bf B302} (1993) 533; 
   {\bf B320} (1994) 400; \\
   D.\ Adams {\it et al.} (SMC), Phys.\ Lett.\ {\bf B329} (1994) 399; \\
   K.\ Abe {\it et al.} (E143), Phys.\ Rev.\ Lett.\ {\bf 74} (1995) 346;
   {\bf 75} (1995) 25.   
   
\bibitem{ellkar}
   J.\ Ellis and M.\ Karliner, Phys.\ Lett.\ {\bf B341} (1995) 397.

\bibitem{adler}
   S.\ L.\ Adler and W.\ A.\ Bardeen, Phys.\ Rev.\ {\bf  182} (1969) 1517;\\ 
   J.\ S.\ Bell and R.\ Jackiw, Nuov.\ Cim.\ {\bf 51A} (1969) 47;\\
   H.\ Fritzsch, M.\ Gell-Mann and H.\ Leutwyler, Phys.\ Lett.\ {\bf B47}
   (1973) 365.

\bibitem{altros}
   A.\ V.\ Efremov and O.\ V.\ Teryaev, Dubna preprint JINR E2--88--287    
   (1988);\\
   G.\ Altarelli and G.\ G.\ Ross, Phys.\ Lett.\ {\bf B212} (1988) 391;\\
   R.\ D.\ Carlitz, J.\ C.\ Collins and A.\ H.\ Mueller, Phys.\ Lett.\ 
   {\bf B214} (1988) 229;\\
   G. Veneziano, Mod.\ Phys.\ Lett.\ {\bf A4} (1989) 1605;\\
   A.\ V.\ Efremov, J.\ Soffer und O.\ V.\ Teryaev,
   Nucl.\ Phys.\ {\bf B346} (1990) 97. 

\bibitem{fri89}
   H.\ Fritzsch, Phys.\ Lett.\ {\bf B229} (1989) 122.

\bibitem{frimin}
   H.\ Fritzsch and P.\ Minkowski, Nuov.\ Cim.\ {\bf 30} (1975) 393.

\bibitem{goltre}
   M.\ L.\ Goldberger and S.\ B.\ Treiman, Phys.\ Rev.\ {\bf 110} (1958) 1178.

\bibitem{rujgla}
   A.\ de Rujula and S.\ L.\ Glashow, Phys.\ Rev.\ Lett.\ {\bf 34} (1975) 46;\\
   A.\ de Rujula, H.\ Georgi and S.\ L.\ Glashow, Phys.\ Rev.\ 
   {\bf D12} (1975) 147;\\
   N.\ Isgur, Phys.\ Rev.\ {\bf D13} (1976) 122.

\bibitem{sca}
   M.\ D.\ Scadron, Phys.\ Rev.\ {\bf D29} (1984) 2076.

\bibitem{kintor}
   R.\ K.\ Carnegie, Phys.\ Lett.\ {\bf B68} (1977) 289; \\
   H.\ J.\ Lipkin, Phys.\ Lett.\ {\bf B72} (1977) 249; \\
   R.\ Kinnunen and N.\ A.\ T\"ornqvist, Lett.\ Nuov.\ Cim.\ {\bf 23} (1978) 
   517;\\
   S.\ Oneda and A.\ Miyazaki, Phys.\ Rev.\ {\bf D40} (1989) 3016; \\
   M.\ Suzuki, Phys.\ Rev.\ {\bf D47} (1993) 1252; \\
   H.\ J.\ Lipkin, Phys.\ Lett.\ {\bf B303} (1993) 119.

\bibitem{bol}
   Mark III Collaboration, T.\ Bolton {\it et al.}, Phys.\ Lett.\ {\bf B278}
   (1992) 495. 

\bibitem{tsai}
   Y.-S.\ Tsai, Phys.\ Rev.\ {\bf D4} (1971) 2821.

\bibitem{bern}
   J.\ Bernstein, {\it Elementary Particles and Their Currents}
   (W.\ H.\ Freeman, 1968).

\bibitem{jaff}
   R.\ L.\ Jaffe, Phys.\ Lett.\ {\bf B193} (1987) 101.

\bibitem{elljaf}
   J.\ Ellis and R.\ L.\ Jaffe, Phys.\ Rev.\ {\bf D9} (1974) 1444.

\bibitem{iofkho}
   B.\ L.\ Ioffe and A.\ Yu.\ Khodjamirian, Sov.\ J.\ Nucl.\ Phys.\ {\bf 55}    
   (1992) 1701.

\bibitem{nashve}
   S.\ Narison, G.\ M.\ Shore and G.\ Veneziano, Nucl.\ Phys.\ {\bf B433} 
   (1995) 209.

\bibitem{belter}
   A.\ V.\ Belitsky, O.\ V.\ Teryaev, Dubna preprint JINR E2--95--374    
   (1995), Phys.\ Lett.\ {\bf B} (to be published).

\bibitem{doming}
   C.\ Dominguez, in preparation.
\end{thebibliography}
\end{document}